# Ethical conceptual replication of visualization research considering sources of methodological bias and practical significance

Ian T. Ruginski*

Department of Geography, University of Zürich

**ABSTRACT**

General design principles for visualization have been relatively well-established based on a combination of cognitive and perceptual theory and empirical evaluations over the past 20 years. To determine how these principles hold up across use contexts and end-users, I argue that we should emphasize conceptual replication focused on determining practical significance and reducing methodological biases. This shift in thinking aims to determine how design principles interact with methodological approaches, laying the groundwork for visualization meta-science.

**Keywords**: visualization, research methods, usability evaluation

**Index Terms**: Human-centered computing—Visualization—Visualization design and evaluation methods

## 1 INTRODUCTION

The past few decades of visualization research have increasingly implemented empirical evaluations [1,2]. As understanding of human perception and decision-making advance, they suggest general principles to guide visualization design. Ample research has been conducted on visual encoding choices, such as color choice, object spacing, and object movement [3,4]. Others have focused on decision-making, noting that specific biases, such as boundary effects when summarizing uncertainty estimates or truncating y-axes, lead to systematic biases and misunderstandings about underlying data distributions [5,6,7].

Along with increases in empirical evaluation come new challenges – such as responsible use of experiments and statistical inference. Methods and statistics training is not always well-integrated with researchers' training, which could inadvertently lead to false, misleading, or overgeneralized claims. This issue is not unique to visualization research and has become a hotly debated issue more broadly in the social sciences, and sometimes referred to as the "replication crisis", resulting in an open science movement [8]. Within visualization research, Kosara and Haroz have called for a revolution focused on replication, which can be achieved by addressing threats to study validity [9].

In this paper, 1) we argue that visualization researchers should familiarize themselves with issues in measurement and statistics, 2) provide examples of where biases can occur, and 3) provide a roadmap for moving forward to address common empirical issues.

## 2 POTENTIAL SOURCES OF BIAS

Sources of bias are inevitable in empirical research, but as responsible researchers, we should take as many steps to reduce bias as reasonably possible. Not only will seeking to minimize biases in our research result in more accurate conclusions with greater external validity and replicability, but will also generate more ethical research by design. I focus on experimental and statistical sources of bias: study sampling bias, and data averaging, modeling, and measurement bias.

* e-mail: ian.ruginski@gmail.com

### 2.1 Sampling bias, ethics, and generalizability

Relying on a majority White, Educated, Industrialized, Rich, and Democratic (WEIRD) sample for research is a well-documented phenomenon in academic research [10]. Not only do biased samples limit the generalizability of studies to minority groups, they also can give misleading estimates of results in the first place. For example, a sample selected based on a collider variable that correlates with two study measures of interest can suggest a correlation in the study sample, even if there is no correlation between the study measures of interest in the general population [11]. This could manifest by selecting for individuals from a third collider variable, such as WEIRD demographics (see Figure 1).

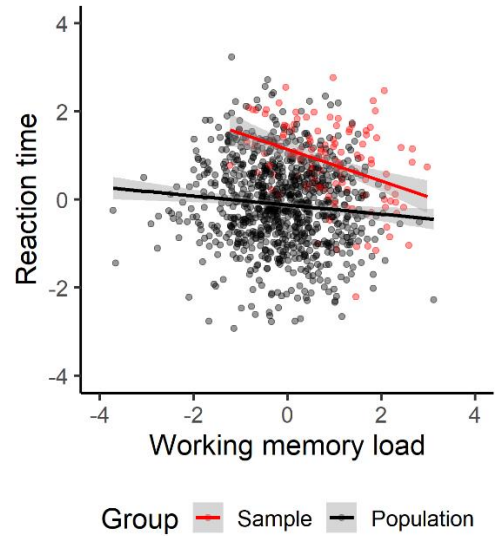

Figure 1: A hypothetical relationship between working memory load and reaction time ($M = 0$, $SD = 1$). Even though there is no relationship between the variables in the population (black dots and line), a spurious correlation might arise in a study due to sampling bias (red dots and line). Code for all figures and statistics available at https://osf.io/ebwx9.

However, in Figure 1 there is no correlation between working memory and reaction time in the general population (simulated population $r = -.01$, 95% $CI = [-.08, .05]$, $n = 1000$), which would have led to false conclusions in this study due to sampling bias (simulated sample $r = -.37$, 95% CI = $[-.51, -.21]$, $n = 125$). This example serves as a call to utilize more generalizable samples and temper claims about the general population from limited samples.

Related, with the COVID-19 pandemic and pervasiveness of the internet, there have been increases in online studies. Visualization researchers have argued for the benefits of Mechanical Turk as a low-cost solution for fast, more diverse, and still accurate research, even replicating Cleveland and McGill's visualization perception results with surprising precision [12]. However, it's easy to overlook the ethical implications of this work: though participants may be willing to complete tasks for 0.01 to .10 cents per task, it does not mean that we should be engaging in labor exploitation. Alternative platforms have emerged which pay participants fair wages, such as Prolific and Gorilla [13, 14]. Shifting to these platforms would result in more ethically conducted research with fair pay, while also potentially reducing sampling bias.

Even when researchers are interested in a subpopulation and fairly compensate participants, results can differ by domain. For example, individuals treat uncertainty in visualizations differently when making real estate decisions versus safety-critical decisions [15]. Data domain and personal background should be considered when making claims from studies, not only visualization design.

## 2.2 Measurement error and parameter bias

Measurement error occurs because we do not have perfect measurements. Even using different rulers to measure distance often results in millimeters of difference in error, due to either the instrument and person. Now imagine that phenomena magnified in the case of measurements of complex human behavior. To address this problem, researchers should validate their measures before implementing them, or use models that address measurement error, such as latent variable modeling [16,17]. Model parameters can also be biased if regression assumptions are violated, or if data is averaged before modeling. One problem due to averaging before modeling occurs when groups in data have differing associations than the averaged data [18] (Simpson's Paradox, see Figure 2).

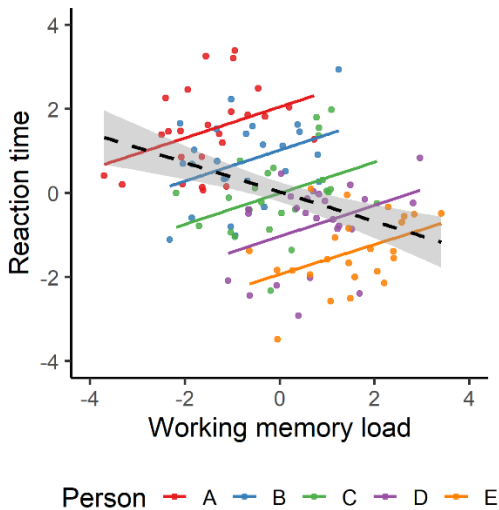

Figure 2: A hypothetical relationship between working memory and reaction time. Even though there is a positive relationship between variables on the within person level (colored dots and lines), there is a negative relationship between variables on the between person level (black dashed line).

To solve this issue, a statistical model should reflect experimental design. For example, an experiment with multiple trials should be modeled in a multilevel model, not averaged, then modeled [19].

Further, meta-insights are sometimes measured to determine the suitability of visualization encodings. For example, users have been asked about trust in uncertainty encodings (e.g. transparency) [20]. This measurement is thought to determine how icons intuitively communicate data concepts to users. However, the visualizations that users prefer are often not the most supportive of decision-making, and in some cases, can be distracting without aiding visualization comprehension [21]. Simlarly to Padilla et al., I argue that we should utilize converging measurements in order to have a more complete picture of decision-making with visualizations, with preferences as one measurement amongst many [22].

## 3 SOLUTIONS AND A WAY FORWARD

### 3.1 Within-participants designs

Within subjects, counterbalanced designs can help reduce sampling bias by manipulating variables such as working memory directly across different trials in the experiment, and presenting multiple trials with different visualizations to a single person. This would help to avoid the averaging and sampling bias problems, with the added bonus of increased statistical power [23]. There are also some drawbacks to the approach, such as the ability of competing visualizations to influence judgments on subsequent trials, as well as the ecological validity of a working memory manipulation during visualization comprehension. Tradeoffs should be weighed based on the practical and theoretical goals of the research.

### 3.2 Towards practical significance

Empirical research often emphasizes statistical significance. For example, we compare whether one group mean is different from another group's mean based on a p-value. A problem with these approaches is that researchers treat statistical inference as a series of binary decisions, with a focus on $p < .05$ as statistically significant. This approach overlooks that p-values can decrease with greater sample size and provide no magnitude information.

These approaches can undermine a key goal of applied research – to determine practical significance. Another approach includes reporting effects of magnitude, such as confidence intervals. However, researchers also often dichotomize these intervals during interpretation (e.g. confidence intervals do not include 0, indicating a significant result) [24]. Solutions include implementing minimal effects hypothesis tests [25], which assess a range of practical significance, and justifying chosen alpha levels [26]. This will encourage consideration of the practical implications of research apriori [27]. Does it matter if there is a 100 millisecond difference in reaction time between visualizations, even if $p < .001$? Kay et al. have also argued for Bayesian approaches as a solution, which address issues of small sample size, integrate past research into models, and generate distributions of effects [28].

### 3.3 Conceptual replication

One way to reframe visualization research is to focus on conceptual replication with practical significance in mind. Exact replication alone is not enough, because exact replication can occur due to correlated measurement error across studies, even if a true effect is not present in the general population [29]. Instead, visualizations should be assessed with a variety of validated measurements, tasks, and domains, systematically determining when changes occur and if these changes are practically meaningful for a given context. While a full discussion of these issues is beyond the scope of this paper, others have considered 1) different contexts when evaluating visualizations [1], and 2) threats to visualization study validity [11].

## 4 CONCLUSION

An increased emphasis on bias and practical significance seeks to accomplish three main goals. First, it will increase generalizability of visualization research to historically overlooked groups. Second, it will help determine how design principles interact with approaches to empirical evaluation, providing a more complete picture of visualization "meta-science." Lastly, it will reduce common methodological pitfalls and threats to validity while providing clearer practical implications.

## 5 ACKNOWLEDGEMENTS

Thank you to the ERC who helped fund this work under grant 740426, and to Steve Haroz who provided a helpful first review.